\documentclass[prd,aps,floats,preprintnumbers,preprint]{revtex4}

\usepackage{amsmath,amssymb,amsfonts}
\usepackage{color,latexsym,placeins}

\usepackage{hyperref}

\newcommand{\be}{\begin{eqnarray}}
\newcommand{\ee}{\end{eqnarray}}
\newcommand{\bea}{\begin{eqnarray}}
\newcommand{\eea}{\end{eqnarray}}

\def\ba{\begin{eqnarray}}
\def\ea{\end{eqnarray}}
\def\be{\begin{equation}}
\def\ee{\end{equation}}

\def\nn{\nonumber}

\renewcommand{\inf}{\infty}
\begin{document}
\title{Newton gauge cosmological perturbations for static spherically symmetric modifications of the de Sitter metric}

\author{Camilo Santa V\'{e}lez$^{2}$ Antonio Enea Romano$^{1,2}$ }
\affiliation
{
$^1${Theoretical Physics Department, CERN, CH-1211 Geneva 23, Switzerland}\\
${}^2$Instituto de Fisica, Universidad de Antioquia, A.A.1226, Medellin, Colombia\\
}



\begin{abstract}
Static coordinates can be convenient to solve the vacuum Einstein's equations in presence of spherical symmetry, but for cosmological applications comoving coordinates are  more suitable to describe an expanding Universe, especially in the framework of cosmological perturbation theory (CPT). 
 Using CPT we develop a  method to transform static spherically symmetric (SSS) modifications of the de Sitter solution from static coordinates to the Newton gauge.  

We  test the method with the Schwarzschild de Sitter (SDS) metric and then derive general expressions for the Bardeen's potentials for a class of SSS metrics obtained by adding to the de Sitter metric a term linear in the mass and proportional to a general function of the radius. Using the gauge invariance of the Bardeen's potentials we then obtain a gauge invariant definition of the turn around radius.

We apply  the method to an SSS solution of the Brans-Dicke theory, confirming the results obtained independently by solving the perturbation  equations in  the Newton gauge. The Bardeen's potentials are then derived for new SSS metrics involving logarithmic, power law and exponential modifications of the de Sitter metric. We also apply the method to SSS metrics which give flat rotation curves,  computing the radial energy density profile in comoving coordinates in presence of a cosmological constant.

\end{abstract}
\maketitle

\section{Introduction}
Spherically symmetric solutions are important because they allow to compute different observables within the solar system or to study the stability of large scale structures \cite{Pavlidou:2013zha,Tanoglidis:2014lea}. They can consequently provide a very important tool to test modified gravity theories \cite{Bhattacharya:2016vur}. 
In general relativity the most general spherically symmetric vacuum solution of the Einstein's field equations with a cosmological constant, the SDS solution,  can be written in static coordinates \cite{Schleich:2009uj}. This is a consequence of the Birkhoff's theorem, but for a general modified theory of gravity such a coordinate system may not be possible. For a modified gravity theory in fact the most general spherically symmetric solution may not be expressed in static coordinates, but only a subclass of the general solutions. This is the consequence of the fact that the Birkhoff's theorem  may not hold in  modified gravity theories \cite{Faraoni:2010rt}.

In cosmology, comoving coordinates are more convenient to describe an expanding Universe, especially in the framework of CPT, and it is therefore useful to find the coordinate transformation from static to comoving coordinates.
In this paper, using CPT, we develop a  method to transform from static coordinates to the Newton gauge SSS  metrics which are modification of the de Sitter solution and  apply it to different known and new cases.
We first apply the method to the SDS metric, confirming it does indeed allow to compute correctly the Newtonian potential due to a point mass in an expanding Universe. 
Using the gauge invariance of the Bardeen's potentials we  obtain a gauge invariant definition of the turn around radius, checking it is consistent with the result obtained in static coordinates for the SDS metric and for other SSS metrics.  
We then compute the Newtonian potentials for  a spherically symmetric solution of the Brans-Dicke theory in static coordinates, confirming the results obtained independently by solving the Brans-Dicke field equations in  the Newton gauge \cite{Bhattacharya:2016vur}.
The Bardeen's potentials are then derived for new SSS metrics involving logarithmic, power law and exponential modifications of the de Sitter metric. We also apply the method to SSS metrics which give flat rotation curves,  computing the radial energy density profile in comoving coordinates in presence of a cosmological constant. 

\section{Static Coordinates}
Assuming a general spherically symmetric  metric ansatz of the type

\be\label{GRAnsatz}
ds^2=F(T,R)dT^2-H(T,R)dR^2- R^2 d\Omega^2,
\ee
one of the Einstein's equations gives $\partial_T f=0$, implying the existence of the well known static coordinates solution
\be\label{SDSStatic}
F(T,R)=H(T,R)^{-1}=\left(1-\frac{2m}{R}-H^2 R^2\right),
\ee
where  have defined $H^2=\Lambda/3$. Substituting the same general ansatz in the field equation of a different gravity theory the equation $\partial_T f=0$ may not hold anymore, and a general spherically symmetric solution may not be written in static coordinates anymore.
The de-Sitter metric, i.e. the $m=0$ limit of the SDS metric,  can also  be written in isotropic coordinates
\bea
R&=&a(t) r \,\label{deSitterTrans1}\,,\\
T&=&t-\frac{1}{2H}\log(r^2a^2(t)-H^{-2}) \,,\label{deSitterTrans2}\\
ds^2&=&dt^2-a^2(t)(dr^2+r^2d\Omega^2),
\eea
where $a(t)=e^{H t}$ is the scale factor.
These coordinates are called comoving coordinates in cosmology because they are interpreted as the coordinates of the observer comoving with the Hubble flow, and for this reason we will use this terminology in the rest of this paper.
We will use this coordinate transformation from static to comoving coordinates to re-write  the SDS metric far from the Schwarzschild radius, i.e. for $m\ll R $, in terms of cosmological perturbations respect to the FRW background.




\section{Cosmological perturbations in a spherically symmetric space and their relation with static coordinates}
For cosmological applications it is useful to re-write the SSS metrics in comoving coordinates. The the most general scalar perturbations respect to the flat FRW background can be written as  \cite{Mukhanov:1990me}
\bea
 ds^2&=&a^2\Bigl\{(1+2\psi)d\tau^2-2\partial_i \omega d\tau dx^i -\left[(1-2\phi)\delta_{ij}+D_{ij}\chi\right]dx^i dx^j\Bigr\}, \label{CPT}
\eea
where $D_{ij}=\partial_i\partial_j-\frac{1}{3}\delta_{ij}\nabla^2$. 

Assuming spherical symmetry the metric can be written as \cite{Notari:2008gi}
\bea\label{CPTspherical}
ds^2&=&a^2  \Bigl[(1+2\psi)d\tau^2 - \left(1-2\phi+\frac{2}{3}\mathcal{E}\right)dr^2 -2\omega'd\tau dr-\left(1-2\phi-\frac{1}{3}\mathcal{E}\right)r^2 d\Omega^2 
\Bigr],
\eea
where
\bea
\mathcal{E}=\chi''-\frac{\chi'}{r} \label{epsilon} \, ,
\eea
and the prime $^{'}$ denotes derivative respect to $r$.

Static coordinates are not commonly used in cosmology, where the comoving coordinates are normally preferred, but can be useful to solve different problems such as  the estimation of the maximum size of gravitationally bounded structures. This is due to the fact that in static coordinates it is more convenient to define the turn around radius \cite{Bhattacharya:2016vur}. For this reason it can be interesting to understand what cosmological perturbations metrics can be written in static coordinates.
In general it is difficult to establish if static coordinates exist for a generic  metric written in the form in eq.(\ref{CPTspherical}), so we will approach the problem from the opposite direction and look for an answer to this question : what form does an SSS metric take in terms of cosmological perturbations theory?

\section{SDS metric in the Newton gauge}
The SDS metric in static coordinates is:
\be
ds^2=\left(1-\frac{2m}{R}-H^2R^2\right)dT^2-\left(1-\frac{2m}{R}-H^2R^2\right)^{-1}dR^2-R^2 d\Omega^2.
\ee
Our goal is to re-write it as a perturbation of the FRW metric, and we will achieve this by performing the coordinate transformation from static to comoving coordinates given in eq.\eqref{deSitterTrans1} and eq.\eqref{deSitterTrans2}. 

Far away from the Schwarzschild horizon ($m\ll R$) we get:


\be\begin{split}
ds^2&=\left[1-\frac{2 m \left(H^2 r^2 a^2+1\right)}{ a r \left(H^2 r^2 a^2-1\right)^2}\right]dt^2-a^2\left[1+\frac{2 m \left(H^2 r^2 a^2+1\right)}{r a \left(H^2 r^2 a^2-1\right)^2}\right]dr^2\\
&-\left[\frac{8 H m a}{\left(H^2 r^2 a^2-1\right)^2}\right]dtdr-r^2 a^2d\Omega^2,
\end{split}\ee

and after introducing conformal time $d\tau=dt/a(t)$ 
\be\begin{split}\label{NoGauge}
ds^2=a^2&\left\{\left[1-\frac{2 m \left(H^2 r^2 a^2+1\right)}{r a \left(H^2 r^2 a^2-1\right)^2}\right]d\tau^2-\left[1+\frac{2 m \left(H^2 r^2 a^2+1\right)}{r a \left(H^2 r^2 a^2-1\right)^2}\right]dr^2\right.\\
&-\left.\left[\frac{8 H m}{\left(H^2 r^2 a^2-1\right)^2}\right]d\tau dr-r^2 d\Omega^2\right\}\,.
\end{split}\ee

Comparing with equation \eqref{CPTspherical} we obtain 
\bea
\psi&=&-\frac{m \left(H^2 r^2 a^2+1\right)}{r a \left(H^2 r^2 a^2-1\right)^2} \,,\label{pertNG}\\
\phi&=&-\frac{H^2 m r^2 a^2+m}{3 r a \left(H^2 r^2 a^2-1\right)^2} \,,\label{pertNG1}\\  
\omega'&=&\frac{4 H m}{\left(H^2 r^2 a^2-1\right)^2} \,,\label{wSDS}\\
\mathcal{E}&=&\frac{2 \left(H^2 m r^2 a^2+m\right)}{r a \left(H^2 r^2 a^2-1\right)^2}.\label{pertNG2}
\eea
After integrating eq.\eqref{epsilon} and eq.\eqref{wSDS} we finally get the scalar  cosmological perturbations variables as defined in eq.(\ref{CPT})
\bea
\omega&=&\frac{2 m \tanh ^{-1}(H r a)}{a}-\frac{2 H m r}{H^2 r^2 a^2-1} \,, \\
\chi&=&\frac{2 m r \left(H r a \tanh ^{-1}(H r a)-1\right)}{a}+\frac{1}{2}r^2 C(\tau)+D(\tau) \,,
\eea
where $C$ and $D$ are functional constants of integration. Since we are only interested in perturbations which should vanish in a limit in which the mass vanishes, the physically  interesting solutions  correspond to  $C=D=0$.

Using CPT we can  derive explicitly the gauge transformation between the static coordinates and the Newton gauge. 
Under an infinitesimal space-time translation of the form
\bea
\tilde{x}^0&=&x^0+\zeta,\\
\tilde{x}^i&=&x^i+\partial^i \beta,
\eea
the gauge transformations are \cite{Notari:2008gi}
\bea\label{gaugeT}
\tilde{\phi}&=&\phi-\frac{1}{3}\nabla^2\beta+\frac{a_\tau}{a}\zeta, \\ 
\tilde{\omega}&=&\omega+\zeta+\beta_\tau,\label{gaugeTw}\\
\tilde{\psi}&=&\psi-\zeta_\tau-\frac{a_\tau}{a}\zeta,\label{gaugeTpsi}\\
\tilde{\chi}&=&\chi+2\beta\,,\label{gaugeTx}
\eea
where we are denoting with a subscript  the derivative respect to conformal time, i.e. for example ${a}_\tau=\frac{da}{d\tau}$.
Imposing the Newton gauge condition 
\be\label{NewtonG}
\omega_N=\chi_N=0,
\ee
after solving the differential equations \eqref{gaugeTw} and \eqref{gaugeTx} we get 
\bea
\label{Trans1}
\beta_N &=&m r \left[\frac{1}{a}-H r \tanh ^{-1}(H r a)\right],\\
\label{Trans2}
\zeta_N &=&\frac{m \left[-r a_\tau-2 H^2 r^2 a^3 \tanh ^{-1}(H r a)+2 H r a^2+2 a \tanh ^{-1}(H r a)\right]}{a^2 \left(H^2 r^2 a^2-1\right)}.
\eea

After substituting eq.(\ref{pertNG}-\ref{pertNG1}) and eq.(\ref{Trans1}-\ref{Trans2}) in the gauge transformations in eq.\eqref{gaugeT} and eq.\eqref{gaugeTpsi} we finally  obtain the perturbations in the Newton gauge 
\bea\label{PsiN}
\Psi_N&=&\psi-\partial_\tau\zeta_N-\frac{a_\tau}{a}\zeta_N=-\frac{m}{a r},\\
\label{PhiN}
\Phi_N&=&\phi-\frac{1}{3}\nabla^2\beta_N+\frac{a_\tau}{a}\zeta_N=-\frac{m}{a r}.
\eea
Instead of finding the transformation taking to the Newton gauge given in eq.(\ref{Trans1}-\ref{Trans2}) we could have also computed the Bardeen's potentials \cite{Bardeen:1980kt} directly from eqs.(\ref{pertNG}-\ref{pertNG2})  
\bea\label{BardeenPsi}
\Psi_B=\psi-\frac{1}{a}\left[a\left(\frac{\chi_\tau}{2}-\omega\right)\right]_\tau &=-\frac{m}{a r},\\
\label{BardeenPhi}
\Phi_B=\phi+\frac{1}{6}\nabla^2\chi-\frac{a_\tau}{a}\left(\omega-\frac{\chi_\tau}{2}\right)&=-\frac{m}{a r}\,.
\eea
As expected the the Bardeen's potentials reduce to the Newton gauge potentials obtained in eq.\eqref{PhiN} and eq.\eqref{PsiN}, and the metric takes the  form
\bea\label{cosmoSDS}
ds^2&=&a^2\left[\left(1-\frac{2m}{a r}\right)d\tau^2 -\left(1+\frac{2m}{a r}\right)\left(dr^2+r^2d\Omega^2\right)\right].
\eea
The metric above  is the weak field limit of the McVittie solutions, and an exact  exact coordinate transformation from the SDS metric  metric in static coordinates to  the McVittie solution is known as shown in Appendix A. We can deduce that the  combination of the radial coordinate transformation $R=a\, r$ and the  gauge transformation given in eq.(\ref{Trans1}-\ref{Trans2}) is the weak field limit of such an exact coordinate transformation.
This is a confirmation that our method is correct, since we have independently obtained the relation between the two metrics. 

For solutions where an exact coordinate transformation is not known, the method we developed has the advantage to allow to write an SSS metric as a perturbed FRW solution even in absence of an exact coordinate transformation.

\subsection{Interpretation of the Newton gauge metric and general applications}
Let's check if the obtained metric is consistent with the expectations of the effects of a point-like source in an expanding Universe. 
The form of the metric in the Newtonian gauge in eq.(\ref{cosmoSDS}) is what one would expect intuitively, since the Newtonian gravitational potential is inversely proportional to the physical distance $R=ar$. We can also observe that the condition $\Phi_N=\Psi_N$ is in agreement with the cosmological perturbations equations. In general relativity in fact the first order cosmological perturbations equations for any isotropic energy-momentum tensor imply $\Phi_N=\Psi_N$, which is clearly also the case for a vacuum solution such as the SDS. 

Note that the method we have adopted allows to obtain the solution of the cosmological perturbations equations in any gauge directly from the SDS metric, without the need to solve the Einstein's equations. This can be useful when studying problems with spherical symmetry such as the calculation of the turn around radius, as we will show later. The correspondence between  static coordinates  and the weak field limit  cosmological perturbations  can also be useful  to study  the non-pertubative regime, which can be for example important for  the Vainshtein mechanism \cite{Vainshtein:1972sx,Babichev:2013usa}.

\section{Weak field limit of SSS metrics as perturbations of FRW}
In a modified gravity theory (MGT) the spherically symmetric vacuum solution associated to a point mass may differ from the SDS metric, and we could make an ansatz of this type 
\bea\label{AnsatzModif}
ds^2&=&\left(1-m h_t(R)-H^2R^2\right)dT^2 -\left(1-m h_r(R)-H^2R^2\right)^{-1}dR^2-R^2d\Omega^2\,,
\eea
where we are not assuming anymore that $g_{tt}=g_{rr}^{-1}$ because in a MGT the field equations may not imply this relation under the assumption of spherical symmetry.
In a generic MGT spherical symmetry may also not imply that, as in GR, $\partial_t g_{tt}=\partial_t g_{rr}=0$, but here we will only consider  solutions which can be written in static coordinates. We will not assume any specific MGT, and adopt a purely phenomenological  approach in order to obtain the Newtonian gauge form of these SSS metrics. These can then be used to test them with  observational data, and only after the metrics compatible with observations have been identified we could try to find which MGT they are solutions of.

After applying to eq.\eqref{AnsatzModif} the coordinate transformations given in eq.\eqref{deSitterTrans1} and  eq.\eqref{deSitterTrans2}, far from the Schwarzschild horizon, i.e. assuming $m\ll R$, we get
\be\label{CosmoModif}\begin{split}
ds^2=a^2 &\left\{\frac{H^6 r^6 a^6-H^4 r^4 a^4 [m h_r(r a)+3]-1+m h_t(r a)}{\left(H^2 r^2 a^2-1\right)^3}d\tau^2\right.\\
&+\frac{ H^2 r^2 a^2 \left[m^2 h_r(r a)^2+mh_r(r a)-mh_t(r a)\right]+3H^2 r^2 a^2}{\left(H^2 r^2 a^2-1\right)^3}d\tau^2\\
&-\left[1+\frac{m h_r(r a) \left(H^2 r^2 a^2-m h_r(r a)-1\right)+H^2 m r^2 a^2 \left(H^2 r^2 a^2-1\right) h_t(r a)}{\left(H^2   r^2 a^2-1\right)^3}\right]dr^2\\
&\left.-\frac{2 H m r a \left[h_r(r a) \left(H^2 r^2 a^2-m h_r(r a)-1\right)+\left(H^2 r^2 a^2-1\right) h_t(r a)\right]}{\left(H^2 r^2 a^2-1\right)^3}d\tau dr-r^2 d\Omega^2\right\}\,.
\end{split}\ee
Comparing the metrics in eq.\eqref{CosmoModif} and eq.\eqref{CPTspherical} we can identify  the perturbation variables in the case of spherical symmetry as
\bea
\psi&=&\frac{ H^2 r^2 a^2m \left[h_r(r a) \left(-H^2 r^2 a^2+m h_r(r a)+1\right)-h_t(r a)\right]+mh_t(r a)}{2 \left(H^2 r^2 a^2-1\right)^3}\,, \\
\phi &=& \frac{m^2 h_r(r a)^2-m\left(H^2 r^2 a^2-1\right) \left[H^2 r^2 a^2 h_t(r a)+h_r(r a)\right]}{6 \left(H^2 r^2  a^2-1\right)^3}\,,\\
\mathcal{E}&=&\frac{m\left(H^2 r^2 a^2-1\right) \left[H^2 r^2 a^2 h_t(r a)+h_r(r a)\right]-m^2 h_r(r a)^2}{\left(H^2 r^2 a^2-1\right)^3} \,, \\
\omega^\prime&=&\frac{H m r a \left(H^2 r^2 a^2-1\right) [h_r(r a)+h_t(r a)]-H m^2 r a  h_r(r a)^2}{\left(H^2 r^2 a^2-1\right)^3}.\label{wGen}
\eea
Solving equation \eqref{epsilon} and integrating eq.\eqref{wGen} we can finally find the perturbations in the general form 
\bea
\omega &=&H m a \int \frac{r [h_r(r a)+h_t(r a)]}{\left(H^2 r^2 a^2-1\right)^2} \, dr,\\
\chi&=&m \int k_1 \int \frac{H^2 k_2^2 a^2 h_t(k_2 a)+h_r(k_2 a)}{k_2 \left(H^2 k_2^2 a^2-1\right)^2} \, dk_2 \, dk_1\\
&&+\frac{1}{2}r^2C +D,\nn
\eea
where $C$ and $D$ are integration constants. Well behaved perturbations require $C=D=0$. These can be replaced in eq.\eqref{BardeenPsi} and eq.\eqref{BardeenPhi} to obtain the Bardeen's potentials.

For  applications such as the study of gravitationally bounded objects we are interested in regions of space-time far from the cosmological horizon, i.e. $a\,r \ll 1/H$ .
Under this assumption the perturbations can be written as
\bea
\psi&=&\-\frac{1}{2} m h_t(r a)\,, \\
\phi &=&-\frac{1}{6} m h_r(r a) [m h_r(r a)+1]\,,\\
\omega&=&H m a \int r [h_r(r a)+h_t(r a)] \, dr \,, \\
\chi&=&m \int k_1 \int \frac{h_r(k_2 a)}{k_2} \, dk_2 \, dk_1.
\eea
In this limit the Bardeen's potentials take the form
\begin{equation}\label{PhiModif}
\Phi_B=\frac{m}{2}\int \frac{h_r(a r)dr}{r}	\,,
\end{equation}
\begin{equation}\label{PsiModif}
\Psi_B=-\frac{m}{2}h_t(a r).
\end{equation}
It is easy to check that for $h_t(R)=h_r(R)=2/R$ the Bardeen's potentials in eq.\eqref{PhiModif} and eq.\eqref{PsiModif} reduce to the SDS Newton gauge perturbations obtained in eqs.\eqref{PhiN} and \eqref{PsiN}.

As shown above, contrary to the case of general relativity, the two potentials $\Phi_B$ and $\Psi_B$ can be different, which is a consequence of the fact that for a general SSS metric  $h_t(R)$ and $h_r(R)$ can be different.   
The Bardeen's potentials can be used to test these SSS metrics using physical observables which are more conveniently computed in the framework of cosmological perturbation theory.
Once the SSS metrics in  agreement with observational data have been identified by using both their static coordinates and cosmological perturbations form, it will be possible to search for the modified gravity theories they are solutions of. The advantage of this approach is that it is independent of the of the modified gravity theory and allows to narrow the search of modified gravity theories to the ones which admit as solutions the SSS metric compatible with observational data.


\section{Gauge invariant definition of the turn around radius}
The turn around radius is the critical distance from the center of a spherically symmetric structure where the radial acceleration vanishes \cite{Pavlidou:2013zha,Tanoglidis:2014lea}.
It can be used as an estimate of the maximum size of gravitationally stable structures and can be  an important observational probe to test the effects of the modification of gravity \cite{Bhattacharya:2015iha,Bhattacharya:2015chc}  or to constrain dark energy \cite{Pavlidou:2014aia}.

The calculation of the turn around radius is more convenient in static coordinates, since for a generic metric of the form in eq.\eqref{GRAnsatz} the radial geodesic equation is 
 \bea
 \frac{d^2 R}{d s^2}&=&\frac{1}{2}H(t,R)\dot{t}^2\frac{\partial F(t,R)}{\partial R}+\frac{\dot{R}^2}{2H(t,R)}\frac{\partial H(t,R)}{\partial R} -\frac{\dot{R}\dot{t}}{H(t,R)}\frac{\partial H(t,R)}{\partial t},
 \eea
which for a static observer defined by $\dot{R}=0$ reduces to
 \bea\label{TARStatic}
 \frac{d^2 R}{d s^2}=\frac{1}{2}H(t,R)\dot{t}^2\frac{\partial F(t,R)}{\partial R} \,,
 \eea
where the dot denotes a derivative respect to the affine parameter $s$. The turn around radius corresponds to the solution of the equation:
\begin{equation}\label{TARStatic}
\partial_R F(R_{TA})=0,
\end{equation}
which in the case of the SDS metric, i.e. when $F(R)=1-2m/R-H^2R^2$, gives
 \be\label{TARFisico}
 R_{TA}=\sqrt[3]{\frac{m}{H^2}} \,.
 \ee
The calculation of the turn around radius in the Newton gauge  \cite{Faraoni:2015zqa} gives instead
\bea\label{CondNewton}
\ddot{a}r-\frac{\Psi_N'}{a}=0 \,.
\eea
After substituting in eq.\eqref{CondNewton} the $\Psi_N$ we obtained in eq.\eqref{PsiN} we obtain the comoving turn around radius:
\be
r_{TA}=\sqrt[3]{\frac{m}{\ddot{a}a^2}},
\ee
which in the SDS case, when $a=e^{Ht}$, gives
\bea
r_{TA}=e^{-Ht}\sqrt[3]{\frac{m}{H^2}}.
\eea
From this equation we  can immediately verify that the physical  radius  $R_{TA}=a \, r_{TA}$ is the same as the one obtained in static coordinates in eq.\eqref{TARFisico}.

Since the turn around radius is an observable quantity it should be gauge-invariant and  we can re-write eq.\eqref{CondNewton} in terms of the Bardeen's potential, to get a gauge invariant  condition:
 \bea\label{CondBardeen}
 \ddot{a}r-\frac{\Psi_B'}{a}=0.
 \eea
 
\section{Gauge invariant computation of the turn around radius for SSS metrics}

The advantage of the gauge invariant definition given in eq.\eqref{CondBardeen} is that we can obtain the turn around radius from the metric of cosmological perturbations in any gauge.
For example starting from the SDS metric written in a gauge different form the Newton gauge, such as in eq.\eqref{NoGauge}, we could compute the Bardeen's potentials defined in eq.\eqref{BardeenPhi} and eq.\eqref{BardeenPsi}, and then solve eq.\eqref{CondBardeen}.

The equivalence between eq.\eqref{TARStatic} and eq.\eqref{CondBardeen} can be shown for  example for the class of SSS metric in eq.(\ref{AnsatzModif}) for which
\begin{equation}
F(R)=1-mh_t(R)-H^2R^2\,,
\end{equation}
and for which the corresponding  Bardeen's potential $\Psi_B$ is given eq.\eqref{PsiModif}. 

Combining eq.\eqref{PsiModif} and eq.\eqref{CondBardeen} we get the general gauge invariant condition for the turn around radius 
\begin{equation}\label{combination}
2H^2 R+m h_t'(R)=0\,
\end{equation}
which is in agreement with eq.\eqref{TARStatic}.

We can apply this method for example to  the SDS metric, corresponding to $h_t=2/R$, for which eq.\eqref{combination} takes the form
\begin{equation}
2H^2 R-2\frac{m}{R^2}=0,
\end{equation}
which gives the solution 
\be 
R_{TA}=\sqrt[3]{\frac{m}{H^2}} \, 
\ee
in agreement with the result obtained in static coordinates.

\section{Newton gauge form for different SSS metrics}
\subsection{ Brans-Dicke Theory}
In Brans-Dicke (BD) theory  the Jebsen-Birkhoff theorem \cite{Faraoni:2010rt} is valid if the scalar field is time independent. As a consequence, under the the assumption of a static scalar field, the static ansatz for the metric adopted in \cite{Bhattacharya:2016vur} should also give the most general spherically symmetric  solution.
Applying a perturbative approach the solution of the field equations can be written as \cite{Bhattacharya:2015iha}
\be
ds^2=\left[1-(1+\epsilon)\frac{2m}{R}-(1-2\epsilon)H^2 R^2\right]dt^2-\left[1-(1-\epsilon)\frac{2m}{R}-(1-4\epsilon) H^2 R^2\right]dR^2-R^2d\Omega^2\,, \label{ds2BD}
\ee
where $\epsilon=\frac{1}{2\omega+3}$. This solution reduces to SDS assuming the observer is far from the cosmological horizon ($R\ll 1/H$) in the limit $\epsilon\to 0$, which is also the limit in which the BD theory reduced to GR. 

Applying the coordinate transformation given in eq.\eqref{deSitterTrans1} and eq.\eqref{deSitterTrans2} and using  conformal time $dt=a \, d\tau$ the metric in eq.\eqref{ds2BD} takes the form
\begin{equation}\begin{split}
ds^2=a^2&\left[\left(1-\frac{2m(1+\epsilon)}{a r}+2\epsilon H^2r^2a^2\right)\tau^2-\left(1-\frac{2m(1+\epsilon)}{a r}-4\epsilon H^2r^2a^2\right)dr^2\right.\\
&\left.-8Hm\,d\tau dr -r^2d\Omega^2\right]\,.
\end{split}
\end{equation}
Comparing with equation \eqref{CPTspherical} and integrating eq.\eqref{epsilon} we get
\bea
\psi&=&\frac{\epsilon H^2 a^3 r^3-m(1+\epsilon)}{a r}, \label{psiBD}\\
\phi&=&\frac{2\epsilon H^2 a^3 r^3-m(1-\epsilon)}{3a r},\label{phiBD}\\
\omega&=&4Hmr, \label{omBD}\\	
\chi&=&-\frac{2mr(1-\epsilon)}{a} -\frac{\epsilon}{2}Ha^2 r^4\label{chiBD}.
\eea
From eq.\eqref{NewtonG} we can find the transformation to go to the Newton Gauge gauge, defined by
\bea
\beta_N&=&-\frac{mr(1-\epsilon)}{a} -\frac{\epsilon}{4}Ha^2 r^4,\label{Trans1BD}\\
\zeta_N&=&-Hmr(3+\epsilon)-\frac{\epsilon}{2}H^3r^4a^3.\label{Trans2BD}
\eea
After substituting eq.(\ref{psiBD}-\ref{phiBD}) and eq.(\ref{Trans1BD}-\ref{Trans2BD}) in the gauge transformations in eq.\eqref{gaugeT} and eq.\eqref{gaugeTpsi} we finally  obtain the perturbations in the Newton gauge 

\bea
\label{PsiBD}\Psi_N&=&\psi-\partial_\tau\zeta_N-\frac{a_\tau}{a}\zeta_N=H^2R^2\epsilon-\frac{m(1+\epsilon)}{R},\\
\Phi_N&=&\phi-\frac{1}{3}\nabla^2\beta_N+\frac{a_\tau}{a}\zeta_N=-H^2R^2\epsilon-\frac{m(1-\epsilon)}{R}\,,
\eea
where $R=a\, r$.
Alternatively we can  obtain the same result without computing any gauge transformation, taking advantage the gauge invariance of the Bardeen's potentials, substituting in eq.\eqref{BardeenPsi} and eq.\eqref{BardeenPhi} the perturbations obtained in eq.(\ref{psiBD}-\ref{chiBD}), getting again
\bea
\label{PsiBD}\Psi_B&=&H^2R^2\epsilon-\frac{m(1+\epsilon)}{R},\\
\Phi_B&=&-H^2R^2\epsilon-\frac{m(1-\epsilon)}{R},
\eea
which coincide with the Newton gauge result as expected, due to the gauge invariance of $\Psi_B$ and $\Phi_B$.

The same result can also be obtained from eq.\eqref{PhiModif} and eq.\eqref{PsiModif} with $h_t=2(1+\epsilon)/R-2\epsilon H^2 R^2/m$ and $h_r=2(1-\epsilon)/R-4\epsilon H^2 R^2/m$.
 The potentials reduce to  the GR result $\Phi_N=\Psi_N=-\frac{m}{a r}$ in eq.(\ref{PsiN},\ref{PhiN}) when $\epsilon\to 0$.

From the Bardeen's potential computed in  eq.\eqref{PsiBD} we get the turn around radius 
\be
r_{TA}=\frac{1+\epsilon}{a}\sqrt[3]{\frac{m}{H^2}}\,,
\ee
which corresponds to the physical radius
\be
R_{TA}=a r_{TA}=\sqrt[3]{\frac{m}{H^2}}(1+\epsilon)\,,
\ee
in agreement with \cite{Bhattacharya:2016vur}.

In general relativity the absence of anisotropic pressure perturbations in the vacuum implies that $\Phi_N=\Psi_N$ while in BD theory the field equations do not imply this anymore, and they can be different.
Note we have  recovered, far form the cosmological horizon, the metric computed in \cite{Bhattacharya:2016vur} solving the perturbation equations in the Newton gauge. This shows explicitly what is the coordinate transformation between the solution in static and comoving coordinates, and that the solutions are indeed the same. The advantage of this approach is that it allows to derive the metric as a Newton gauge perturbation of the FRW solution directly from the metric in static coordinates, without the need to solve again the perturbed  field equations as it was done in \cite{Bhattacharya:2016vur}.

\subsection{Power law modifications of the de Sitter metric}
Let's consider the  sub-class of SSS metrics given in eq.(\ref{AnsatzModif}) corresponding to this choice of $h_t,h_r$
\bea
h_t(R)&=&\lambda_1 R^{n_1} \,,\\ 
h_r(R)&=&\lambda_2 R^{n_2}. 
\eea
Following the same procedure shown in the previous section we first identify the perturbations in the spherically symmetric form given in  eq.(\ref{CPTspherical})
\bea
\phi&=&-\frac{1}{6}\left[m \lambda_2 (r a)^n_2+m^2 \lambda_2^2 (r a)^{2n_2}\right] \,, \\
\psi&=&-\frac{1}{2}m \lambda_1 (r a)^{n_1} \,, \\
\omega&=&Hma r^2\left[\frac{\lambda_1(a r)^{n_1}}{2+n_1}+\frac{\lambda_2(a r)^{n_2}}{2+n_2}\right] \,, \\
\chi&=&\frac{\lambda_2 m r^2 (a r)^{n_2} \left(\frac{\lambda_2 m (r a)^{n_2}}{n_2+1}+\frac{4}{n_2+2}\right)}{4 n_2} \,.
\eea
We can then compute  the Bardeen's potentials in the region $m \ll R \ll 1/H$
\bea
\Psi_B&=-\frac{m\lambda_1 (ra)^{n_1}}{2},\\
\Phi_B&=\frac{m\lambda_2 (r a)^{n_2}}{2n_2}.
\eea
The difference between the $\Psi$ and $\Phi$ is due to the difference between $g_{tt}$ and $g_{rr}^{-1}$ in static coordinates, and  it could  arise in vacuum solutions of modified gravity theories admitting this SSS solution. 

The turn around radius in comoving coordinates is given by
\begin{equation}
r_{TA}=\frac{1}{a}\left(-\frac{m \lambda_1 n_1}{2H^2}\right)^{\frac{1}{2-n_1}},
\end{equation}
while in  static coordinates coordinates is
\begin{equation}
R_{TA}=\left(-\frac{m \lambda_1 n_1}{2H^2}\right)^{\frac{1}{2-n_1}},
\end{equation}
which reduces to eq.\eqref{TARFisico} when $\lambda_1=2$ and $n_1=-1$.
\subsection{Exponential modifications of the de Sitter metric}
For the  sub-class of SSS metrics given in eq.(\ref{AnsatzModif}) corresponding to this choice of $h_t,h_r$
\bea
h_t(R)&=&\lambda_1 e^{b_1 R} \,, \\
h_r(R)&=& \lambda_2 e^{b_2 R} \,,
\eea
for the metric perturbations we get
\bea
\phi&=&-\frac{m \lambda_2 e^{b_2 r a}+m^2 \lambda_2^2 e^{2b_2 r a}}{6 },\\
\psi&=&-\frac{m \lambda_1 e^{b_1 r a}}{2 },\\
\omega&=&\frac{H m \left[b_1^2 \lambda_2 e^{b_2 r a} (b_2 r a-1)+b_2^2 \lambda_1 e^{b_1 r a}(b_1 r a-1)\right]}{2 b_1^2 b_2^2 a},\\
\chi&=&\frac{\lambda_2^2 m^2 e^{2 b_2 r a}}{8 b_2^2 a^2}+\frac{\lambda_2 m e^{b_2 r a}}{2 b_2^2 a^2}+\frac{1}{2} \lambda_2^2 m^2 r^2 \text{Ei}(2 b_2 r a)+\frac{1}{2} \lambda_2 m r^2 \text{Ei}(b_2 r a)\\
&-&\frac{\lambda_2^2 m^2 r e^{2b_2 r a}}{4 b_2 a}-\frac{\lambda_2 m r e^{b_2 r a}}{2 b_2 a},
\eea
where $Ei(z)$ is the exponential integral function defined as
\be
Ei(z)=-\int_{-z}^\inf \frac{e^{-t}}{t}dt.
\ee
The corresponding Bardeen's  potentials are
\bea
\Psi_B&=&-\frac{1}{2} \lambda_1 m e^{b_1 r a},\\
\Phi_B&=&\frac{1}{2} \lambda_2 m \text{Ei}(b_2 r a).
\eea
Note that in the above expressions we are only giving the leading order terms in the region $m \ll R \ll 1/H$.

\subsection{Logarithmic modifications of the de Sitter metric}
In the sub-class of SSS metrics given in eq.(\ref{AnsatzModif}) corresponding to this choice of $h_t,h_r$
\bea
h_t(R)&=&\lambda_1 \log{b_1 R} \,, \\  
h_r(R)&=&\lambda_2\log{b_2 R}, 
\eea
the cosmological perturbations are 
\bea
\phi&=&-\frac{m \lambda_2  \log \left(a r b_2\right)+m^2 \lambda_2^2  [\log \left(a r b_2\right)]^2}{6},\\
\psi&=&-\frac{m \lambda_1  \log \left(a r b_1\right)}{2 },\\
\omega&=&\frac{1}{4} H m r^2 a [2 \lambda_1 \log (b_1 r a)+2 \lambda_2 \log (b_2 r a)-\lambda_1-\lambda_2],\\
\chi&=&\frac{1}{6} \lambda_2^2 m^2 r^2 \log ^3(b_2 r a)-\frac{1}{4} \lambda_2^2 m^2 r^2 \log ^2(b_2 r a)+\frac{1}{4} \lambda_2^2 m^2 r^2 \log (b_2 r a)\\
&+&\frac{1}{4} \lambda_2 m r^2 \log ^2(b_2 r a)-\frac{1}{4} \lambda_2 m r^2 \log (b_2 r a)-\frac{1}{8} \lambda_2^2 m^2 r^2+\frac{1}{8} \lambda_2 m r^2.
\eea
and the corresponding Bardeen's  potentials far away from the horizons ($m\ll R\ll 1/H$) are
\bea\label{PsiNLog}
\Psi_B&=&-\frac{1}{2} \lambda_1 m \log \left(b_1 r a \right),\\
\Phi_B&=&\frac{1}{4} \lambda_2 m \left[\log \left(b_2 r a \right)\right]^2.
\eea
The turnaround radius can be calculated analytically by solving eq.\eqref{CondBardeen}, obtaining 
\begin{equation}
r_{TA}=\frac{1}{H a}\sqrt{\frac{m\lambda_1}{2}}\,,
\end{equation}
which in static coordinates gives
\begin{equation}
R_{TA}=\frac{1}{H}\sqrt{\frac{m\lambda_1}{2}}.
\end{equation}

\section{SSS metrics giving flat rotation curves}
The SSS metric\cite{Boehmer:2007kx} which gives flat rotation curves is
\begin{equation}\label{ds2rot}
ds^2=\left(\frac{R}{R_c}\right)^{2v^2}dt^2-\left[1-v^2 f(R)-H^2R^2\right]dR^2-R^2d\Omega^2.
\end{equation}

Applying the coordinate transformation given in eq.\eqref{deSitterTrans1} and eq.\eqref{deSitterTrans2} and using conformal time, in the low tangential velocity regime and far from the cosmological horizon ($v\ll 1, R\ll 1/H$), the metric takes the following form 
\begin{equation}\begin{split}
ds^2=a^2 &\left\{\left[1+H^2r^2a^2+2v^2\log\frac{a r}{R_c}\right]d\tau^2-\Bigg[1-2H^2 r^2 a^2-v^2 f(a r)\Bigg]dr^2\right.\\
&\left.-\left[6H^3r^3a^3+2Hrv^2a\left(f(a r)+2 \log\frac{a r}{R_c}\right)\right]dr d\tau-r^2 d\Omega^2\right\}\,.
\end{split}\end{equation}

Comparing with equation \eqref{CPTspherical} and integrating eq.\eqref{epsilon}  we get 
\bea
\psi&=&\frac{H^2r^2a^2}{2}+v^2\log\left(\frac{a r}{R_c}\right)\,,\label{rotpsi}\\
\phi&=&\frac{1}{3}H^2r^2a^2+\frac{v^2}{6}f(a r)\,,\\
\omega&=&\frac{1}{2} \int_0^r \left\{-2 H r v^2 a \left[f(r a)+2 \log \left(\frac{r a}{R_c}\right)\right]-6 H^3 r^3 a^3\right\} \, dr\,,\\	
\chi&=&\int_0^r k_1 \int_0^{k_1} \left(-\frac{v^2 f(k_2 a)}{k_2}-2 H^2 k_2 a^2\right) \, dk_2 \, dk_1\,.\label{rotchi}
\eea
We can then compute the Bardeen's potential by substituting eqs.(\ref{rotpsi}-\ref{rotchi}) in eq.\eqref{BardeenPsi}
\begin{equation}
\Psi_B=v^2\log \left(\frac{R}{R_c}\right)-\frac{1}{2}H^2R^2.
\end{equation}
This agrees with the results obtained in \cite{Boehmer:2007kx} with an additional term due to the cosmological constant. 

In order to better understand this solution we can compute the radial energy density profile using the Einstein's equation $G^0_{\ 0} = 8\pi T^0_{\ 0}=8\pi \rho$
\begin{equation}
8\pi\rho= G^0_{\ 0}=3H^2 +2\nabla^2\Phi-6H\Phi^\prime=\frac{2v^2}{a^2 r^2}-6H^2.
\end{equation}
In terms of the physical radius $R=a\,r$  we get
\begin{equation}
\rho(R)=\frac{2v^2}{8\pi R^2}-\frac{3H^2}{4\pi},
\end{equation}
and assuming flatness the total energy contained inside a sphere of radius $R$ is obtained by integrating the density $\rho$
\begin{equation}
M(R)=\int_0^{R} 4\pi R'^2\rho(R') dR'= \int_0^{R}  \left(v^2-3H^2 R'^2\right)dR'=v^2 R-H^2 R^3.
\end{equation}
We have obtained the expected linear behavior as in \cite{Boehmer:2007kx} with an additional cosmological constant term.
This confirms  that the method we have used to re-write the SSS solution in eq.(\ref{ds2rot}) as a perturbed FRW metric is giving  correct results.
\section{Conclusions}

Using cosmological perturbation theory  we have developed a  method to transform from static coordinates to the Newton gauge  spherically symmetric metrics  which are modifications of the de Sitter metric. We  applied the method to the Schwarzschild de Sitter  metric, confirming it does indeed allow to compute correctly the Newtonian potential due to a point mass in an expanding Universe. Using the gauge invariance of the Bardeen's potentials we have  obtained a gauge invariant definition of the turn around radius, checking it is consistent with the result obtained in static coordinates for the SDS metric.

We have then applied the method to derive general expressions for the Bardeen's potentials for a class of SSS metrics obtained by adding to de Sitter metric a term linear in the mass and proportional to a general function of the radius. 
We have  computed the Bardeen's potentials for  a SSS solution of the Brans-Dicke theory  in static coordinates, confirming the results obtained independently by solving the Brans-Dicke field equations in  the Newton gauge.
The Bardeen's potentials have also been derived explicitly for logarithmic, power law and exponential modifications. 
Finally  we applied the  method to study SSS metrics which give flat rotation curves, and after re-writing them as perturbations of a FRW background  we have computed the energy density radial profile, obtaining the expected behavior together with a contribution from  the cosmological constant.



The method we have developed could be used to study new types of SSS metrics and to asses their compatibility with observations using physical quantities which are more conveniently computed in the framework of cosmological perturbations theory. In this way the search for modified gravity theories could be narrowed to those which admit as solutions the SSS metrics which better fit observational data. 

\acknowledgments
We thank Constantinos Skordis for useful discussions.

\appendix

\section{Exact coordinate transformation for the SDS metric in the Newton gauge}
It is easy to check that the McVittie's metric \cite{McVittie:1933zz}
\bea
ds^2&=&\left(\frac{1-\frac{m}{2a r}}{1+\frac{m}{2a r}}\right)^2 d\tilde{t}^2 -a^2\left(1+\frac{m}{2a r}\right)^4\left[dr^2+r^2d\Omega^2\right],
\eea
in the weak field limit $m \ll r$ coincides with the metric obtained in eq.\eqref{cosmoSDS}. This hints to the fact that the metric we obtained is the weak field limit of the McVittie metric, which is also the weak field limit of the SDS metric in comoving coordinates. This implies that there should exist an exact coordinate transformation between the McVitties metric and the SDS metric in static coordinates. It is in fact known that the McVittie's metric can be obtained from the SDS metric by the following coordinate transformation \cite{Gao:2004wn},\cite{Kaloper:2010ec} 
\bea\label{MCVT}
t&=&\tilde{t}+\gamma(R),\\
\label{MCVR}R&=&e^{H\tilde{t}}r+m+\frac{m^2}{4e^{H\tilde{t}}r},
\eea
where $\gamma(r)$ is defined by:
\bea\label{MCVG}
\frac{d\gamma}{dR}=-\frac{HR^2}{\sqrt{R-m}(1-\frac{m}{R}-H^2R^2)},
\eea
and  $a(\tilde{t})=e^{H \tilde{t}}$.
We can conclude that the gauge transformation in eqs.\eqref{Trans1},\eqref{Trans2}, which was obtained using cosmological perturbation theory, is the perturbative limit of the transformation given in eqs.\eqref{MCVT}-\eqref{MCVG}.

For the SDS metric the exact coordinate transformation between static coordinates and the Newton gauge is known but for a general static spherically symmetric metric it may be more difficult to find it, while the gauge transformation approach can be always adopted, and as we have  verified in different ways, it gives the correct Newton gauge form of the SDS metric. 

\bibliography{Bibliography}
\bibliographystyle{h-physrev4}

\end{document}